\documentclass{article}

\font\bBB=msbm8 at 9pt
\font\bB=msbm10 at 11pt
\def\bBL{\mbox{\bB L}}
\def\bBC{\mbox{\bB C}}
\def\bBR{\mbox{\bB R}}
\def\bBBL{\mbox{\bBB L}}

\def\la{\langle\,}
\def\ra{\,\rangle}
\def\ee{\end{equation}}
\def\dcE{{\cal E}}
\def\tr{{\rm Tr}\,}
\def\1{{\mathchoice{\rm 1\mskip-4mu l}{\rm 1\mskip-4mu l}%
{\rm 1\mskip-4.5mu l}{\rm 1\mskip-5mu l}}}
\newtheorem{defn}{Definition}
\newtheorem{thm}{Theorem}
\newtheorem{rem}{Remark}

\title{\bf On a Stroboscopic Approach to Quantum Tomography of Qudits
Governed by Gaussian Semigroups}
\author{Andrzej Jamio{\l}kowski \\[2ex]
Institute of Physics \\ Nicolaus Copernicus University \\
87--100 Toru\'n, Poland}
\date{}

\begin{document}

\maketitle

\begin{abstract}
In this paper, we discuss the minimal number $\eta$ of
observables $Q_1,\ldots,Q_\eta$, where expectation values at some
time instants $t_1,\ldots,t_r$ determine the trajectory of a
$d$-level quantum system (``qudit") governed by the Gaussian
semigroup
\[
\Phi(t)\rho\;=\;\frac{1}{\sqrt{2\pi
t}}\int\limits_{-\infty}^{\infty}ds\, e^{-s^2/(2t)}e^{-iHs}\rho
e^{iHs}.
\]
We assume that the macroscopic information about the system in
question is given by the mean values
$\dcE_j(Q_i)=\tr(Q_i\rho(t_j))$ of $n$ selfadjoint operators
$Q_1,\ldots,Q_n$ at some time instants $t_1<t_2<\ldots<t_r$, where
$n<d^2-1$ and $r\leq {\rm deg}\,\mu(\lambda,\bBBL)$. Here
$\mu(\lambda,\bBBL)$ stands for the minimal polynomial of the
generator
$$
\bBBL\rho\;=\;-\frac12 \bigl[H,[H,\rho]\bigr]
$$
of the Gaussian flow $\Phi(t)$.
\end{abstract}

\section{Introduction}
\setcounter{equation}{0}

The ability to create, manipulate and characterize quantum states
is becoming  increasingly important in physical research with
implications for other areas of science, such as: quantum
information theory, quantum communication and computing. According
to one of the basic assumptions of quantum mechanics, the
achievable
 information about the state of a physical system is encoded in the
density matrix $\rho$, which allows one to evaluate all possible
expectation values of observables trough the formula
\begin{equation}\label{aj1.1}
\la Q\ra \;=\;\tr(\rho Q)\,,
\end{equation}
where $Q$ is a self-adjoint operator representing a particular
physical quantity.

Thus, in order to have full information about a given quantum
system we need to know its density matrix $\rho$. In particular, a
very useful tool in this regard is quantum state tomography (QST)
which provides means for the complete reconstruction of the
density matrix for a qudit (or a set of qubits). The general
procedure relies on the ability to reproduce a large number of
identical states and to perform a series  of measurements on
complementary aspects of the state within an ensemble.

Suppose that we can prepare a quantum system repeatedly in the
same state and  make a series of experiments such that we can
measure the expectation values
\begin{equation}\label{aj1.2}
 \dcE_t(Q)\;=\;\tr(Q\rho(t))
\end{equation}
of some observables $Q_1,\ldots,Q_n$ at different time moments
$t_1<t_2<\ldots<t_r$. The fundamental question arises:
\par\medskip\noindent\it
Can we find the average value of any desired operator from the set
of measured outcomes of a given set $Q_1,\ldots,Q_n$
\begin{equation}\label{aj1.3}
  \left[\begin{array}{ccc}
    \dcE_{t_1}(Q_1) & \cdots & \dcE_{t_r}(Q_1) \\
    \vdots & \cdots & \vdots \\
     \dcE_{t_1}(Q_n) & \cdots & \dcE_{t_r}(Q_n)
  \end{array}\right],
\end{equation}
where $0\leq t_1<\ldots<t_r\leq T$, for an interval $[0,T]$?
\par\medskip\rm
Among the existing tomographic techniques for quantum systems, the
so-called homodyne tomography has received much attention  in the
literature \cite{aj1,aj2,aj3}. In the phase-space formulation of
quantum mechanics there is a one-to-one relation between a quantum
state and the so-called Wigner function. Its marginals are
accessible experimentally, and an inverse (Radon) transformation
allows one to reconstruct the phase-space distribution associated
with the unknown quantum state.

The question of how to reconstruct states of $d$-level systems
(qudits) is also natural. In this case various methods have been
proposed to determine $\rho$ \cite{aj4,aj5,aj6}. If the problem
under consideration is static, then the state of a $d$-level open
quantum system (a qudit) can be uniquely determined only if
$n=d^2-1$ expectation values of linearly independent observables
are at our disposal. However, if we assume that we know the
dynamics of our system, i.e.~we know the generator of the time
evolution, then we can use the stroboscopic approach based on
(\ref{aj1.3}). In general, the term ``tomography"  will be used
collectively to denote any kind of state-reconstruction method.

With reference to the terminology used in the system theory, we
assume the following definition:
\begin{defn}\rm
A $d$-level open quantum system ${\cal S}$ is said to be
$(Q_1,\ldots,Q_n)$-reconstructible on an interval $[0,T]$, if
there exists at least one set of time instants $0\le
t_1<\ldots<t_r\leq T$ such that the state trajectory can be
uniquely  determined by the correspondence
\begin{equation}\label{aj1.4}
  [0,T]\ni
  t_j\;\longmapsto\;\dcE_{t_j}=\tr(\rho(t_j)Q_i)
\end{equation}
for $i=1,\ldots,n$, $j=1,\ldots,r$.
\end{defn}
The above definition is equivalent to the following one
\begin{defn}\rm
A $d$-level open quantum system ${\cal S}$ is said to be
$(Q_1,\ldots,Q_n)$-reconstructible on an interval $[0,T]$, if for
every two trajectories with distinct initial states there exists
at least one  $\hat t\in[0,T]$ and at least one operator
$Q_k\in\{Q_1,\ldots,Q_n\}$ such that
\begin{equation}\label{aj1.5}
  \tr(Q_k\rho_1(\hat t))\;\neq\;\tr(Q_k\rho_2(\hat t))\,.
\end{equation}
\end{defn}
\begin{rem}\rm
In the above definitions we assume that the time evolution of the
system  is given in terms of a completely positive semigroup of
operators. Arguments in favour of completely positive semigroups
as the foundation of non-Hamiltonian dynamics as well as the
discussion of  properties of such semigroups can be found in
papers of Kraus~\cite{aj7}, Lindblad~\cite{aj8}, and Gorini {\it
et al.\/}~\cite{aj9}.

In  particular, in Lindblad's paper \cite{aj8} the general form of
the generator of an arbitrary completely positive semigroup was
derived. A linear operator $\bBL$ on a set of linear operators
$B({\cal H}):=M(\bBC^d)$, where ${\cal H}\simeq\bBC^d$ is a
$d$-dimensional Hilbert space and $M$ denotes the set of matricies
with complex entries, proves to be the generator of a completely
positive semigroup if and only if it can be represented in the
form
\begin{equation}\label{aj1.6}
  \bBL\rho\;=\;-i[H,\rho]+\frac12\sum_{j=1}^\kappa\Big([V_j\rho,V_j^*]+
  [V_j,\rho V_j^*]\Big)\,,
\end{equation}
where $V_j\in B({\cal H})$ for $j=1,\ldots,\kappa$, and $H$ is a
self-adjoint operator also belonging to $B({\cal H})$ (cf.~also
\cite{aj10}).
\end{rem}
\begin{rem}\rm
It is important that for the number $r$ of time instants
$t_1,\ldots,t_r$ we do not formulate any restriction (except that
it is finite).
\end{rem}
\begin{rem}\rm
The question of obvious physical interest is to find the minimal
number of observables $Q_1,\ldots,Q_\eta$ for which the $d$-level
quantum system ${\cal S}$ with the generator $\bBL$ can be
$(Q_1,\ldots,Q_\eta)$-reconstructible. It can be shown that if the
time evolution of the system ${\cal S}$ is described by the master
equation,
\begin{equation}\label{aj1.7}
  \frac{d}{dt}\rho(t)\;=\;\bBL\rho(t)\,,
\end{equation}
then there exists \cite{aj4,aj5} a set of observables
$Q_1,\ldots,Q_\eta$, where
\begin{equation}\label{aj1.8}
  \eta\;=\;\max_{\lambda\in\sigma(\bBBL)}\Big\{{\rm dim}\,{\rm
  Ker}\,(\lambda\1-\bBL)\Big\}
\end{equation}
such that the system ${\cal S}$ is
$(Q_1,\ldots,Q_\eta)$-reconstructible. Moreover, if we have
another set of observables
$\widetilde{Q}_1,\ldots,\widetilde{Q}_{\tilde\eta}$
with this property,
 then $\tilde\eta\geq\eta$. The number $\eta$
given by (\ref{aj1.8}) we will call {\it the index of cyclicity\/}
of the system ${\cal S}$.
\end{rem}

\section{Polynomial Representation of Flows}
\setcounter{equation}{0}

The main idea of the stroboscopic approach to quantum tomography
is based on the polynomial representation of the flow defined by
the general master equation. Namely, we have
\begin{equation}\label{aj2.1}
\Phi(t)\;=\;\exp(\bBL t)\;=\;\sum_{k=0}^{m-1}\alpha_k(t)\bBL^k\,,
\ee
where by Cauchy's theorem
\begin{equation}\label{aj2.2}
\alpha_k(t)\;:=\;\frac1{2\pi i}\oint_{\partial D}\frac{\mu_k(z)}
{\mu(z,\bBL)}\exp(tz)\,dz\,. \ee In the above expression $\partial
D$ is any simple closed contour enclosing the spectrum of the
operator $\bBL$ in the complex plane and
\begin{equation}\label{aj2.3}
\mu(z,\bBL)\;=\;\sum_{k=0}^{m-1}d_k z^k \ee denotes the minimal
polynomial of the generator $\bBL$. It is interesting that there
are ways to compute the functions $\alpha_k(t)$ in (\ref{aj2.1})
without summing the exponential series or without knowing the
Jordan canonical form of $\bBL$. Namely,  differentiating
(\ref{aj2.1}) with respect to $t$ and using the minimal polynomial
of $\bBL$ one finds that the functions $\alpha_k(t)$ for
$k=0,\ldots,m-1$ satisfy the system of equations
\begin{eqnarray}
\frac{d\alpha_0(t)}{dt} &=& d_0\alpha_{m-1}(t)\,, \nonumber \\
\frac{d\alpha_1(t)}{dt} &=& \alpha_0(t)+d_1\alpha_{m-1}(t)\,,\label{aj2.4} \\
\cdots & & \cdots\qquad\cdots\qquad\cdots \nonumber \\
\frac{d\alpha_{m-1}(t)}{dt} &=&
\alpha_{m-2}(t)+d_{m-1}\alpha_{m-1}(t)\,,\nonumber
\end{eqnarray}
with initial conditions $\alpha_k^{(i)}(0)=\delta_{ik}$. It can be
shown that functions $\alpha_k(t)$ are mutually linearly
independent, therefore for a given $T$ there exists at least one
set of time instants $t_1,\ldots,t_m$ ($m={\rm
deg}\,\mu(\lambda,\bBL^*)$) such that $0\leq t_1<\ldots<t_m\leq T$
and ${\rm det}\,[\alpha_k(t_j)]\neq 0$.

Taking into account the above conditions one finds that the state
$\rho(0)$ can be determined uniquely (and the trajectory
$\Phi(t)\rho(0)$ can be reconstructed) if and only if operators of
the form $(\bBL^*)^k Q_i$ for $i=1,\ldots,n$ and
$k=0,1,\ldots,m-1$ span the space $B({\cal H})$.

If the dynamical semigroup is completely positive, then the
general form of the generator $\bBL$ is given by (\ref{aj1.6}). In
this case the criterion for  reconstructibility of a $d$-level
quantum system can be formulated using the operators $H$ and
$V_j$. In particular, if we consider an isolated quantum system
characterized by a Hamiltonian $H_0$ and $V_j=0$ for
$j=1,\ldots,\kappa$, then the minimal number of observables
$Q_1,\ldots,Q_{\eta}$ for which the system is
$(Q_1,\ldots,Q_{\eta})$-reconstructible is given by
$$
\eta\:=\:n_1^2+n_2^2+\cdots+n_r^2\,,$$ where $n_i=\mbox{\rm
dim\,Ker\,}(\lambda_1I-H_0)$ for all $\lambda_i\in\sigma(H_0)$,
$i=1,\ldots,r$ (for details cf.~\cite{aj5}).

\section{The Minimal Number of Observables for Qudits Governed by Gaussian Semigroups}
\setcounter{equation}{0}

Let us assume that the time evolution of a $d$-level quantum
system ${\cal S}$ is described by the generator $\bBL$ given by
\begin{equation}\label{aj3.1}
\bBL\rho\;=\;\frac12\bigl\{[H\rho,H]+[H,\rho
H]\bigl\}\;=\;-\frac12\bigl[H,[H,\rho]\bigl] \ee that is, the
semigroup $\Phi(t)$ has the form (cf.\ e.g.~\cite{aj11})
\begin{equation}\label{aj3.2}
\Phi(t)\rho\;=\;\frac{1}{\sqrt{2\pi
t}}\int\limits_{-\infty}^{\infty}ds\, e^{-s^2/(2t)}e^{-iHs}\rho
e^{iHs}. \ee The symbol $H$ in (\ref{aj3.1}) and (\ref{aj3.2})
denotes a self-adjoint operator with the spectrum
\begin{equation}\label{ak3.3}
\sigma(H)\;=\;\{\lambda_1,\ldots,\lambda_m\}\,. \ee In the sequel
$n_i$  stands for the multiplicity of the eigenvalue $\lambda_i$
for $i=1,\ldots,m$. One can assume that the elements of the
spectrum of $H$ are numbered in such a way that the inequalities
$\lambda_1<\lambda_2<\ldots<\lambda_m$ are fulfilled. The
following theorem holds:
\begin{thm}\label{th1}
The index of cyclicity of the Gaussian semigroup with a generator
$\bBL$ given by (\ref{aj3.1}) is expressed by the formula
\begin{equation}\label{aj3.4}
\eta\;=\;\max\{\kappa,\gamma_1,\ldots,\gamma_r\}\,, \ee where
$r=(m-1)/2$ if $m$ is odd or $r=(m-2)/2$ if $m$ is even, and
\begin{eqnarray}
\kappa &:=& n_1^2+n_2^2+\ldots+n_m^2\,, \\
\gamma_k &:=& 2\sum_{i=1}^{m-k}n_i\,n_{i+k}\,.
\end{eqnarray}
\end{thm}
\par\medskip\noindent
{\it Proof.}\quad
In order to determine the value of $\eta$ for the generator $\bBL$
defined by (\ref{aj3.1}) we must find the number of nontrivial
invariant factors of the operator $\bBL$. Let us observe that if
$\sigma(H)=\{\lambda_1,\ldots,\lambda_m\}$ then the spectrum of
the operator $\bBL$ is given by
\begin{equation}\label{aj3.7}
\sigma(\bBL)\;=\;\Big\{\nu_{ij}\in\bBR\,;\;\nu_{ij}=(\lambda_i-\lambda_j)^2\,,\;i,j=1,\ldots,m\,\Big\}\,.
\ee
The above statement follows from the fact that the operator $\bBL$ can also be represented as
\begin{equation}\label{aj3.8}
\bBL\;=\;H^2\otimes\1 + \1\otimes H^2-2 H\otimes H\,, \ee where
$\1$ denotes the identity in the space ${\cal B}({\cal H})$. Since
$H$ is self-adjoint therefore the algebraic multiplicity of
$\lambda_i$, i.e.~the multiplicity of $\lambda_i$ as the root of
the characteristic polynomial of $H$, is equal to the geometric
multiplicity of $\lambda_i$, $ n_i={\rm dim}\,{\rm
Ker}\,(\lambda_i\1-H)$\,. Of course, we have $n_1+\ldots+n_m={\rm
dim}\,{\cal H}$.

From (\ref{aj3.8}) we can see that the multiplicities of the
eigenvalues of the operator $\bBL$ are not determined uniquely by
the multiplicities of $\lambda_i\in\sigma(H)$. But if we assume
that $\lambda_1<\ldots<\lambda_m$ and
$\lambda_k=(k-1)c+\lambda_1$, where $k=1,\ldots,m$, and $c={\rm
const}>0$, then the multiplicities of all eigenvalues of $\bBL$
are given by
\begin{equation}\label{aj3.9}
\gamma_{|i-j|}\;=\;{\rm dim}\,{\rm Ker}\,[(\lambda_i-\lambda_j)^2\1-\bBL]
\ee
for $i\neq j$ and
\begin{equation}\label{aj3.10}
{\rm dim}\,{\rm Ker}\,(\bBL)\;=\;n_1^2+\ldots+n_m^2\;=\;\kappa \ee
when $i=j$. Now, as we know, the minimal number of observables
$Q_1,\ldots,Q_\eta$ for which the qudit $\cal S$ can be
$(Q_1,\ldots, Q_\eta)$-reconstructible is given by (\ref{aj1.8}),
so in our case
\begin{equation}\label{aj3.11}
\eta\;=\;\max_{i,j=1,\ldots,m}\Big\{{\rm dim}\,{\rm
Ker}\,[(\lambda_i-\lambda_j)^2\1-\bBL]\Big\}\,, \ee where
$\lambda_i\in\sigma(H)$. Using the above formulae and the
inequality $\gamma_k<\kappa$ for $k>r$, where $r$ is given by
$(m-1)/2$ if $m$ is odd and $(m-2)/2$ if $m$ is even, we can
observe that also without the assumption
$\lambda_k=(k-1)c+\lambda_1$ one obtains
\begin{equation}\label{aj3.12}
\eta\;=\;\max\{\kappa,\gamma_1,\ldots,\gamma_r\}\,.
\ee
This completes the proof.
\par\medskip\noindent
According to Theorem~\ref{th1} if the quantum system governed by a
Gaussian semigroup is $(Q_1,\ldots, Q_n)$-reconstructible then the
number $n$ of observables must satisfy the inequality $n\ge\eta$.
In this case there exists a set of time instants
 $t_1<t_2<\ldots<t_m$ ($m={\rm
deg}\,\mu(\lambda,\bBL)$) such that the knowledge of the
expectation values $\dcE_j(Q_i)=\tr(\rho(t_j)Q_i)$ for
$i=1,\ldots,n$ and $j=1,\ldots,m$ uniquely determines the
trajectory of the system.

The natural question arises: what are the criteria governing the
choice of time instants $t_1,\ldots,t_m$? The following theorem
holds:
\begin{thm}\label{th2}
Let us assume that $0\leq t_1<t_2<\ldots<t_m\leq T$. Suppose that
the mutual distribution of time instants $t_1,\ldots,t_m$ is
fixed, i.e.~a set of nonnegative numbers $c_1<\ldots<c_m$ is given
and $t_j:=c_jt$ for $j=1,\ldots,m$, and $t\in\bBR_+$\,. Then for
 $T>0$ the set
    $$\tau(T):=\Bigl\{(t_1,\ldots,t_m):~~t_j=c_jt,~~0\le t\le \frac{T}{c_m} \Bigr\}$$
 contains almost all
sequences of time instants $t_1,\ldots,t_m$, i.e.~all of them
except a finite number.
\end{thm}
\par\medskip\noindent
{\it Proof.}\quad
As one can check, the expectation values $\dcE_{t_j}(Q_i)$ and the
operators $(\bBL^*)^kQ_1i$ are related by the equality
\begin{equation}\label{aj3.13}
\dcE_{t_j}(Q_i)\;=\;\sum_{k=0}^{m-1}\alpha_k(c_jt)
\Big((\bBL^*)^kQ_i,\rho_0\Big)\,, \ee where we assume that
$t_j=c_jt$ and the bracket $(\cdot,\cdot)$ denotes the
Hilbert-Schmidt product in ${\cal B}({\cal H})$. One can determine
$\rho_0$ from (\ref{aj3.13}) for all those values $t\in\bBR_+$ for
which the determinant $\alpha(t)$ is different from zero, i.e.
\begin{equation}\label{aj3.14}
\alpha(t)\;:=\;{\rm det}\,[\alpha_k(c_jt)]\;\neq\;0\,. \ee One can
prove that the range of the parameter $t\in\bBR_+$ for which
$\alpha(t)=0$ consists only of isolated points on the semiaxis
$\bBR_+$, i.e.~does not possess any accumulation points on
$\bBR_+$. To this end let us note that since the functions
$t\to\alpha_k(t)$ for $k=0,1,\ldots,m-1$, are analytic on $\bBR$,
the determinant $\alpha(t)$ defined by (\ref{aj3.14}) is also an
analytic function of $t\in\bBR$. If $\alpha(t)$ can be proved to
be nonvanishing identically on $\bBR$, then, making use of its
analyticity, we shall be in position to conclude that the values
of $t$, for which $\alpha(t)=0$, are isolated points on the axis
$\bBR$.

It is easy to check that for $k=m(m-1)/2$
\begin{equation}\label{aj3.15}
\frac{d^k\alpha(t)}{dt^k}\Big|_{t=0}=\prod_{1\leq j<i\leq
m}(c_i-c_j)\,.
 \ee
According to the assumption $c_1<c_2<\ldots<c_m$, we have
$\alpha^{(k)}(0)\neq 0$ if $k=m(m-1)/2$. This means that the
analytic function $t\to\alpha(t)$ does not vanish identically on
$\bBR$ and the set of values of $t$ for which $\alpha(t)=0$ cannot
contain accumulation points. In other words, if we limit ourselves
to an arbitrary finite interval $[0,T]$, then $\alpha(t)$ can
vanish only on a finite number of points belonging to $[0,T]$.
This completes the proof.

\end{document}